\documentclass[epsf,twocolumn,preprintnumbers]{revtex4}

\usepackage{amsmath,amssymb,amsfonts}
\usepackage{graphics}
\usepackage{dcolumn} 
\usepackage{bm,graphicx,color}
\usepackage{epsfig}
\usepackage{comment}
\newcommand{\ttn}{ThTaN$_3$}

\begin{document}
\title{Perovskite ThTaN$_3$: a large-thermopower topological crystalline insulator
}
\author{Myung-Chul Jung$^1$}
\author{Kwan-Woo Lee$^{1,2}$}
\email{mckwan@korea.ac.kr}
\author{Warren E. Pickett$^3$}
\email{pickett@physics.ucdavis.edu}
\affiliation{
 $^1$Department of Applied Physics, Graduate School, Korea University, Sejong 33019, Korea\\
 $^2$Division of Display and Semiconductor Physics, Korea University, Sejong 33019, Korea\\
 $^3$Department of Physics, University of California, Davis, California 95616, USA
}
\date{\today}
\begin{abstract}
ThTaN$_3$, a rare cubic perovskite nitride 
semiconductor, has been studied using {\it ab initio} methods.
Spin-orbit coupling (SOC) results in band inversion and a band gap of 150 meV
at the zone center.
Despite trivial $Z_2$ indices, two pairs of spin-polarized surface bands cross
the gap near the zone center, 
indicating that this system is a topological crystalline insulator 
with the mirror Chern number of $|{\cal C}_m|=2$ protected by the mirror and $C_4$ rotational symmetries. 
Additionally, SOC doubles the Seebeck coefficient,
leading to a maximum of $\sim$400 $\mu$V/K at 150 K for carrier-doping levels of several $10^{17}$/cm$^3$. 
ThTaN$_3$ combines excellent bulk thermopower with parallel conduction
through topological surface states that may point towards new possibilities for platforms for large engineering
devices with ever larger figures of merit. 
\end{abstract}
\maketitle

\begin{figure}[tbp]
{\resizebox{7cm}{4cm}{\includegraphics{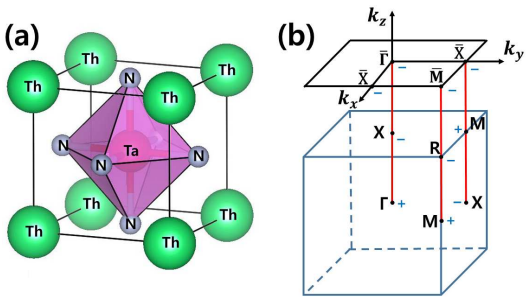}}}
\caption{(a) Crystal structure of ThTaN$_3$,
 plotting using the {\sc vesta} program.\cite{vesta}    
(b) bulk and (001) surface Brillouin zones (BZs) with high symmetry points.
The $\pm$ symbols represent the party of all occupied bands of \ttn~
at each time-reversal invariant momentum (TRIM).
}
\label{str}
\end{figure}

\section{Introduction}
After the discovery of time-reversal $Z_2$ topological insulators (TIs)\cite{kane05},
Fu proposed another type of TI protected by crystal symmetries such as mirror 
or rotational symmetry\cite{tci_fu}. 
This class is called the topological crystalline insulator (TCI),
expected to display exotic quantum phenomena including topological superconductivity.\cite{ando15}
Through model calculations in three-dimensional systems, 
Fu suggested both the $n$-fold rotation $C_n$ ($n$=4,6) and time-reversal symmetries\cite{tci_fu}
could produce TCI, 
whereas Alexandradinata {\it et al.}\cite{alex14} and Alexandradinata and Bernervig\cite{alex16} further proposed that $n$-fold rotation 
and reflection $C_{nv}$ symmetries 
of $n$=3, 4 and 6 may lead to the TCI without help of the time-reversal 
symmetry. 
In contrast to ordinary TIs, 
need for spin-orbit coupling (SOC) is not a requirement for a TCI,
but many prospective TCI materials suggested by first-principles calculations require 
SOC to invert band characters, as in \ttn~ investigated here. 
However, SOC will affect the final gap size, so heavy atoms may still be more
favorable as in conventional TIs.

The combination of large bulk thermopower and gapless TI boundary states represents
a new frontier in topological materials. In finite samples the conductivity 
provided by (topological) surface states impacts
the resulting thermopower as discussed by
Xu {\it et al.}\cite{xu2014,xu2017} The surface conductance leads to an anomalous
Seebeck effect that can be tuned by engineering the size and shape of the sample.
The first TCIs were proposed in IV-VI semiconductors of a simple rock-salt structure
with mirror symmetry\cite{nacl1,nacl2}, such as SnTe, PbTe, PbSe, SnS, and SnSe,
which had been known as good thermoelectric materials\cite{singh10,roy15}.
Soon after these theoretical predictions, the topological character of these 
compounds was observed\cite{tanaka,dziawa,hasan},
specifically, a double Dirac-cone structure through (spin) 
angle-resolved photoemission spectroscopy (ARPES).
In contrast to the ordinary $Z_2$ index class of TIs showing an odd number 
of Dirac points in a surface state,
TCIs have an even number of Dirac points\cite{nacl1}.
Moreover, scanning tunneling microscopy and ARPES studies showed massive Dirac cones 
induced by breaking the symmetry in these TCIs, 
suggesting a possibility for engineering the Dirac band gap\cite{okada,zel,wojek}.
Their two-dimensional monolayer cousins are also suggested to support TCI phases\cite{monoNaCl1,monoNaCl2}.
It has been suggested that the rock-salt structure PbPo shows both TCI and ferroelectric instability\cite{min}.
In additional to the rock-salt compounds, TCI phases have been theoretically predicted in pyrochlore oxides\cite{fiete},
antiperovskite compounds\cite{hsh14}, the orthorhombic perovskite SrIrO$_3$\cite{kee}, 
and full-Heusler compounds\cite{pham}. 
However, exotic physical properties of TCIs have not been established
due to the lack of the combination of both theoretical predictions and 
experimental realization. 

In this Rapid Communication we investigate topological and thermoelectric properties 
of the time-reversal symmetric cubic perovskite nitride \ttn~ 
possessing both the fourfold rotation and mirror symmetries utilized by Fu\cite{tci_fu} 
but now using first-principles approaches.
Nitride perovskites are quite rare compared to the oxides\cite{marq} and with different 
properties, including that itinerant hole or electron doping is expected to be much 
easier in the nitride\cite{ScN-MgO} due to the smaller electronegativity 
and larger polarizability of N relative to O.
The cubic perovskite insulator \ttn, displayed in Fig. \ref{str}(a), 
was obtained using solid-state synthesis methods by Brese and DiSalvo\cite{ttn3_exp},
but characterization has been limited.
Theoretical calculations have confirmed the semiconducting character
and stability of the cubic phase\cite{shein,matar,polfus}.
The results we present indicate that perovskite \ttn~ is not only a TCI, but also displays
highly favorable thermoelectric properties.
 
The organization of the Rapid Communication is as follows. The theoretical methods are 
presented in Sec. II. Sec. III contains the main results on the electronic 
structure, the TCI characteristics, and the carrier density and temperature
dependence of the thermopower. In Sec. IV we provide a brief summary. 

\section{Theoretical Approach}
The experimental diffraction data\cite{ttn3_exp} indicate the cubic perovskite structure, 
and density functional calculations obtain positive elastic constants 
for the cubic structure\cite{shein}. 
Our {\it ab initio} calculations, performed using the experimental lattice parameter
 $a=4.02$ \AA,\cite{ttn3_exp} which is close to our optimized value of $4.06$ \AA, were performed 
using the accurate all-electron full-potential code {\sc wien2k}\cite{wien2k}.
Selected results were confirmed with another all-electron full-potential code {\sc fplo}\cite{fplo}. 
The Perdew-Burke-Ernzerhof generalized gradient approximation (GGA)
was used as the exchange-correlation functional\cite{gga}.
SOC has been included in the results we present except where noted.
The Brillouin zone was sampled with a uniform fine $k$ mesh of $21\times 21\times 21$
to check the narrow gap nature carefully.
In {\sc wien2k}, the basis was determined by $R_{mt}K_{max}=7$ 
and augmented atomic radii in atomic units: Th 2.4, Ta 2.0, and N 1.7. For {\sc fplo}
the default basis functions for the constituent atoms were used.
The second variation precedure \cite{Koelling_80} is used to include SOC effects in {\sc wien2k}.
About 1-Ry width of conduction bands is kept for the second diagonalization, 
giving nearly perfect convergence of the SOC calculation. 

\begin{figure}[tbp]
{\resizebox{8cm}{6cm}{\includegraphics{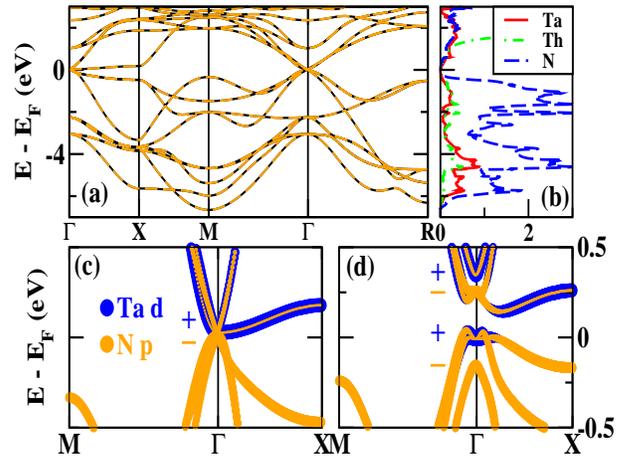}}}
\caption{Top: (a) Nonmagnetic GGA band structure of \ttn~
 in the region containing N $2p$, Ta $5d$, and Th $6d$ and $5f$ orbitals. 
 The nearly flat Th $5f$ bands appear around 3 eV, 
 relative to the Fermi level $E_F$ which is set to zero.
 The orange dashed lines indicate bands obtained from Wannier interpolation,
 showing an excellent representation of the bands.   
 (b) The corresponding atom-projected densities of states (DOSs), 
 in units of states per eV.
 Bottom: enlarged fatband structures of (c) GGA and (d) GGA+SOC 
 along the $M-\Gamma-X$ line near $E_F$. 
 The band inversion induced by SOC at the $\Gamma$ point is evident
 in panel (d). 
 With SOC included, at the $\Gamma$ point the characters of each band are 
 the singlet N $p$, doublet Ta $d$, doublet N $p$, and singlet Ta $d$,
 from the lower to higher energy.
 The $\pm$ symbols at the $\Gamma$ point denote the parity of each band.
}
\label{band}
\end{figure}

To investigate for topological character, our calculated band structure and eigenfunctions were fit 
using the {\sc wannier90} code\cite{wan} with an initial basis of N $2p$, Ta $5d$, Th $6d$, and $5f$ orbitals,
resulting in an excellent fit  in the range --7 eV to 4 eV relative to the Fermi level $E_F$ 
[see Fig. \ref{band}(a)].
From these results, the surface states were calculated through the Green's-function approach 
implemented in the {\sc wanniertools} code\cite{ch_suf}.
The hybrid Wannier charge center is also provided by the {\sc Z2pack} code\cite{z2pack}.

Thermoelectric properties were studied using the {\sc boltztrap} code\cite{boltz}
based on the semiclassical Bloch-Boltzmann transport theory\cite{boltztheory} 
with a constant scattering time
approximation. A much denser regular $k$ mesh containing up to 60000 $k$ points 
was used since these calculations were very sensitive to details of band structure
in the region of the gap.  

\section{Results}

\begin{figure*}[tbp]
{\resizebox{11cm}{11cm}{\includegraphics{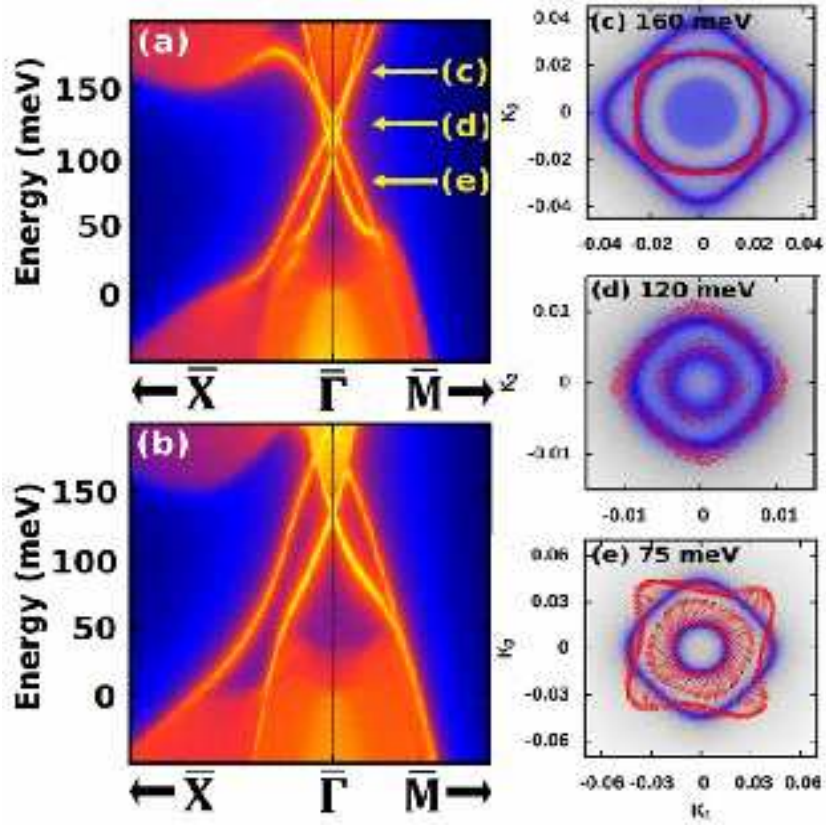}}}
\caption{The (001) surface spectral functions (SOC included) 
obtained from the Green's-function method for (a) ThN and (b) TaN$_2$ terminations. 
The Fermi contours are shown in blue.
Without SOC (not shown here), no surface band appears.
Panels (c)-(e) display the spin texture (shown with red arrows), 
having only in-plane components, 
on the double Dirac cone at energies denoted in panel (a).
The arrows provide the spin direction. Note the changes in scale in panels
(c)-(e).
}
\label{sf}
\end{figure*}

\subsection{Electronic structure}
First we address the electronic structure obtained with a dense $k$ mesh
to investigate the character of its small gap carefully.
The full range GGA band structure and the corresponding atom-resolved DOSs
given in Fig. \ref{band}(a) and \ref{band}(b), respectively,
indicate that this system is a small gap band insulator\cite{gap},
consistent with the formal charge configuration of Ta$^{5+}$, Th$^{4+}$, and N$^{3-}$.
The valence bands have the fully filled N $2p$ character,
whereas the bottom of the conduction bands has Th $t_{2g}$ character 
as shown in Fig. \ref{band}(c) and \ref{band}(d).
Thus \ttn~ shows a common $p-d$ direct gap between two threefold degenerate bands at the zone center.

Now specific effects of SOC are considered.
The bottom panels of Fig. \ref{band} shows enlarged band structures of GGA and GGA+SOC 
overlapped by the fatband coloring showing Ta $5d$ and N $2p$ characters.
Inclusion of SOC leads to splitting the threefold bands on either side 
of the gap into a doublet and singlet and enhancing the energy gap to 150 meV.
The gap is smaller by 10 meV at the optimized volume.
Most importantly, the band splitting results in band inversion at $\Gamma$.
As shown in Fig. \ref{band}(c), before SOC 
the threefold band at the top of the valence band arising from $2p$ states of all
N ions has negative parity,
whereas the Ta $t_{2g}$ triplet at the bottom of the conduction band has positive parity.
SOC results in splitting and inverting the parities: The parity of each band in the doublet 
at the bottom of the conduction band is negative [see Fig. \ref{band} (d)].
This band inversion suggests a nontrivial topological character as will be discussed 
in the next subsection.

\begin{figure}[tbp]
\resizebox{8.5cm}{8.5cm}{\includegraphics{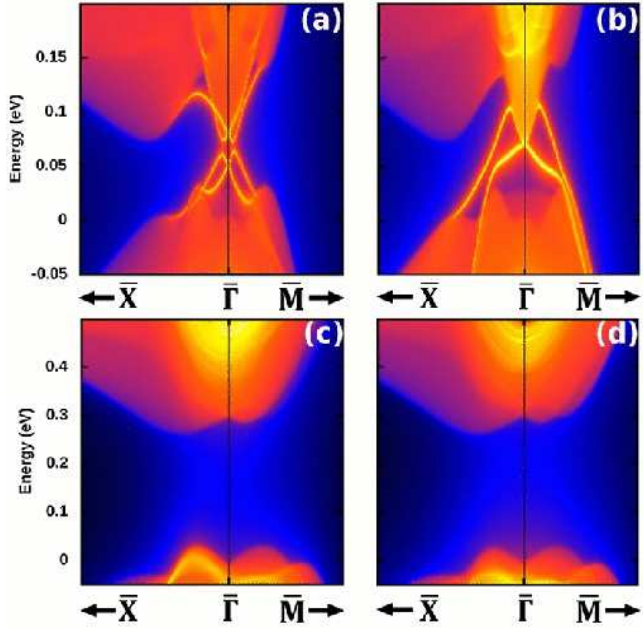}}
\caption{The (001) surface spectral function (SOC included) of ThN (left)
and TaN$_2$ (right) terminations when breaking symmetries.
For the upper panel, the Ta ion is displaced by 0.01 \AA~ in the (100) direction 
to remove only the mirror symmetry.
For the low panel, the ion is displaced by the same amount in the (111) direction
to break both the mirror and rotational symmetries. 
}
\label{bsf}
\end{figure}

\subsection{Topological crystalline insulating phase}
Our calculations of the topological character begin from the Wannier representation,\cite{wan} 
shown in Fig. \ref{band}(a).
The $Z_2$ indices $\nu_0;(\nu_1\nu_2\nu_3)$ are calculated\cite{z2}
from the parties of all occupied bands, excluding the core orbitals, at TRIMs, shown in Fig. \ref{str}(b). 
The indices are $0;(000)$, indicating that \ttn~ is topologically trivial. Specifically,
two negative-parity valence-band states have interchanged with two positive-parity
conduction-band states, and the change in occupied TRIM parities by two leaves vanishing Z$_2$ indices.
To gather further evidence, the hybrid Wannier charge centers are calculated for 
half of the BZ on the mirror plane\cite{z2pack}.
Consistent with the $Z_2$ indices,
the hybrid center plot shows an even number of crossings, indicating that a $Z_2$ TI 
has not emerged (see the Supplemental Material\cite{ttn3_supple}.

Next, we performed surface-state calculations by the Green's-function approach. 
Figure \ref{sf}(a) and \ref{sf}(b) show the surface spectral functions for 
ThN and TaN$_2$ surface terminations,
respectively, using (001) surfaces [given the cubic symmetry, only (001)
needs to be considered].
For both terminations, two pairs of surface bands crossing the gap appear at the zone center,
although distinctions verify that they are inequivalent.

The spin textures, pictured in Fig. \ref{sf}(c)-\ref{sf}(e), 
are calculated in three regimes: below the lower Dirac point,
above the upper Dirac point, and midway between the two points.
At 160 meV in Fig. \ref{sf}(c), the texture on the outer
Fermi contour is clockwise with opposite chirality on the inner contour. 
At 120 meV [the intermediate regime, Fig. \ref{sf}(d)] 
both textures are counterclockwise.
At 75 meV the texture on the outer Fermi contour is counterclockwise
with opposite chirality on the inner contour.

To reveal the origin of this topological character, 
crystalline symmetries have been broken by displacing the Ta ion in two directions.\cite{ttn3_supple}
First, the ion was displaced by 0.01 \AA~ along the (100) direction to remove the
$x\rightarrow -x$  mirror symmetry. 
The upper panel of Fig. \ref{bsf} shows that elimination of this mirror symmetry
destroys the surface state of the ThN termination (the disconnect of bands
near 0.06 eV energy), whereas that of the TaN$_2$ survives.
Second, both the mirror and the rotational symmetries were discarded by shifting Ta 
along the (111) direction by 0.01 \AA.
The bottom panel of Fig. \ref{bsf} reveals that both surface states have disappeared.
This killing of surface bands by breaking of symmetry 
is evidence that \ttn~ is a 
TCI protected by $C_4$ symmetry rather than by mirror symmetry. Similar effects
have been observed in the antiperovskite compounds\cite{hsh14}.
Also, these two pairs of surface bands indicate that
the mirror Chern number ${\cal C}_M=({\cal C}_{+i}-{\cal C}_{-i})/2$ is two, 
where ${\cal C}_{\pm i}$ are individual Chern numbers for Bloch eigenstates 
with eigenvalues $\eta =\pm i$\cite{mirror_chern}.
Note that two Dirac cones in \ttn~ concurrently emerge at $\bar{\Gamma}$,
in contrast to the previous TCIs with $|{\cal C}_M|=2$\cite{nacl1,monoNaCl2},
which have two Dirac cones separately appearing at two different TRIMs.

\begin{figure}[tbp]
{\resizebox{8cm}{8cm}{\includegraphics{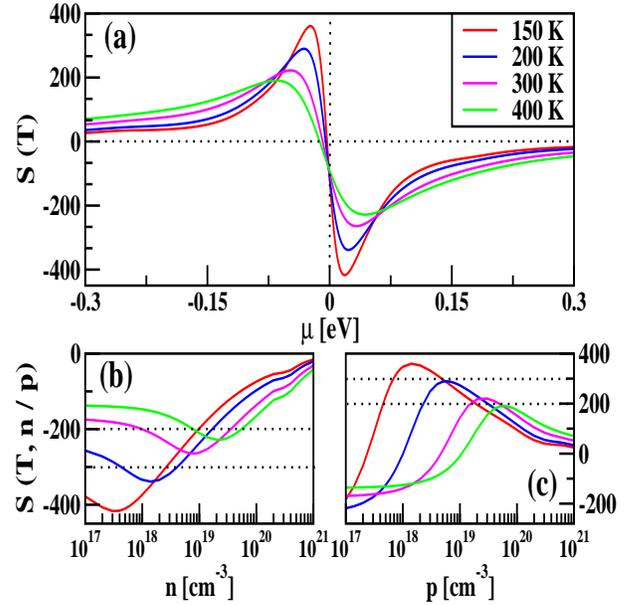}}}
\caption{Seebeck coefficients $S(T)$ (in units of $\mu$V/K)
versus (a) band filling $\mu$, (b) and (c) concentration of doped carriers 
at various temperatures in the range of 150 K -- 400 K; SOC is included.
This behavior is similar to that without SOC (not shown here),
but is roughly doubled by SOC. 
}
\label{seebeck}
\end{figure}

\subsection{Thermoelectric properties}
As mentioned above, the first TCIs were realized in promising thermoelectric compounds.\cite{nacl1,nacl2} 
There is however no direct relation between thermoelectric and topological properties
aside from the commonality of possessing small band gaps.
First, thermoelectric properties are a bulk property, whereas the impact of the topological
character resides in the boundary states.
The Seebeck coefficient is determined solely by the bands $E(\vec{k})$, 
their derivatives, and the Fermi level $E_F$\cite{boltztheory}, 
whereas the topological states are
the result of topological features arising from inversion and mixing of the bulk 
valence and conduction bands. 
Nevertheless, one may simplistically anticipate that a high thermopower is achieved
in TCIs or TIs with a narrow gap since SOC, which is often significant 
in topological matters, leads to considerable changing in dispersion around $E_F$, 
hence may provide the steep DOS that often signals large thermopower.
Such a connection occurs in this system, as we now discuss.

As shown in Fig. \ref{band}(b) and \ref{band}(c),
inclusion of SOC substantially reduces the dispersion of  
the doubly degenerate bands at the top of the valence band 
and the bottom of the conduction band along the $\Gamma-X$ line,
resulting in a steeper DOS that roughly doubles the Seebeck coefficient.
Figure \ref{seebeck} shows $S(T)$ (SOC included) over the range 
150 K -- 400 K in temperature. 
Due to the narrow gap, $S(T)$ rapidly increases in magnitude 
at very small levels of doping and reaches the very large value
of  $\sim$ 400 $\mu$V/K at 150 K.
Upon increasing $T$, the maximum in $S(T)$ monotonically decreases.
However, it retains excellent thermoelectric performance  
(200 $\mu$V/K -- 300 $\mu$V/K)\cite{singh13}, 
indicated by the dashed lines in Fig. \ref{seebeck}, 
over a range of 10$^{17}$/cm$^{-3}$ -- 10$^{19}$/cm$^{-3}$ doping level
({\it i.e}, $\sim$10$^{-4}$ -- 10$^{-6}$ carriers per formula unit) below $T=400$ K.

\section{Summary}

In \ttn, which is a rare cubic nitride perovskite, 
spin-orbit coupling leads to enlarging the gap to 150 meV and inverting valence and conduction bands.
Interestingly, two surface band Dirac cones in the gap concurrently appear at the zone center,
although this system shows trivial $Z_2$ indices and character of the hybrid Wannier charge center.
This even number of surface bands indicates that \ttn~ is a topological crystalline insulator 
with the mirror Chern number $|{\cal C}_m|=2$,
protected by the mirror and fourfold rotational symmetries.
Additionally, due to the narrow gap and less dispersive bands near $E_F$ induced by SOC,
this system shows a very high Seebeck coefficient with a maximum of $\sim$ 400 $\mu$V/K at 150 K,
suggesting a possible application as an element in a thermoelectric device(see the Supplemental Material\cite{ttn3_supple}).
Our findings indicate that stoichiometric \ttn~ is a TCI also with a good thermoelectric properties,
inviting further theoretical and experimental researches. 
Most currently known topological insulators suffer from defects in the bulk 
that degrade the insulating bulk to semiconducting, thus precluding identification of 
and therefore application of the topological surface bands. The sample quality did not allow determination of stoichiometry in \ttn\cite{ttn3_exp}, but this material is chemically distinct from previous TIs and thus provides new opportunities.

\section{Acknowledgments}
We acknowledge Y.-K. Kim for useful discussion of topological insulators and
K.-H. Ahn and Y.-J. Song for useful technical discussions. Discussion on
nitride perovskites with A. P. Ramirez, M. Subramanian, and T. Siegrist
were appreciated.
This research was supported by NRF of Korea Grant No. NRF-2016R1A2B4009579
(M.C.J and K.W.L) and by U.S. DOE BES Grant No. DE-FG02-04ER46111 (W.E.P.).

\newpage
~
\newpage

\begin{center}
\section{Supplemental Material}
\end{center}

\subsection{topological properties}

\begin{figure*}[tbp]
\vskip 8mm
\rotatebox{-90}{\resizebox{7.5cm}{9cm}{\includegraphics{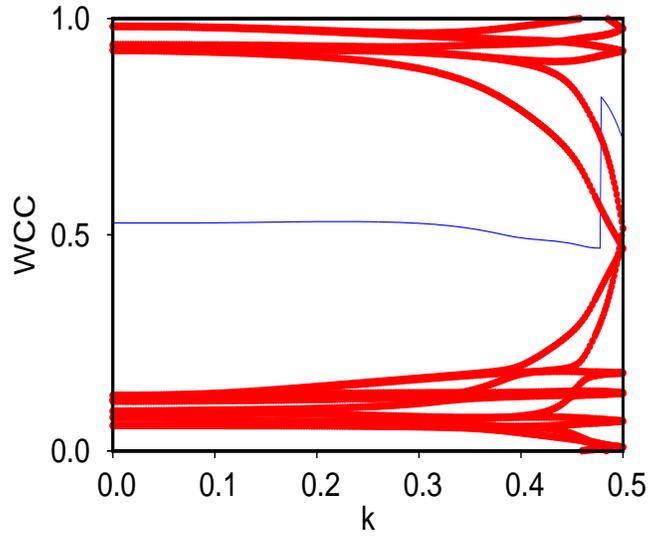}}}
\caption{(Color online) Hybrid Wannier charge center (red, thick lines) plot of \ttn~ 
across half of the Brillounin zone
in the $k_z=0$ plane, showing an even number of crossings between the charge center 
and largest gap function.
 The blue (thin) line denotes largest gap function.
 Here, the wave vector $k$ along the (100) direction is given in unit of $\pi/a$.
}
\label{sfig1}
\end{figure*}

In this brief Supplemental Material file  we provide more evidence 
of the topological crystalline insulator
character of \ttn. One type of evidence involves the number of
hybrid Wannier center crossings (HWCC) across the zone. Figure~\ref{sfig1}
reveals two crossings, an even number representative of a topological
crystalline insulator (TCI). 

\begin{figure*}[tbp]
\resizebox{16cm}{10cm}{\includegraphics{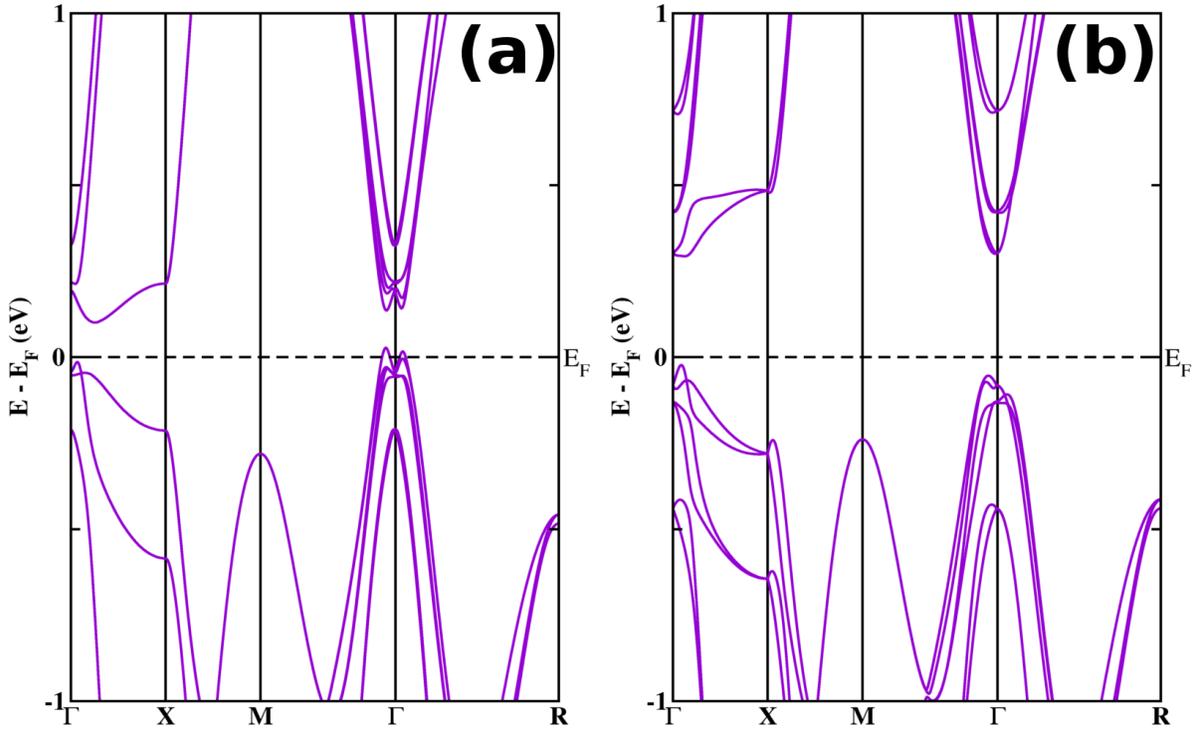}}
\caption{(Color online) Enlarged GGA+SOC band structures, near the Fermi energy $E_F$, 
for breaking (a) only mirror and (b) both mirror and rotational symmetries.
}
\label{sfig2}
\end{figure*}

It was noted in the main text that destruction of mirror or 4-fold rotation
symmetries destroyed the TCI character. In Fig.~\ref{sfig2} the band structures 
are displayed after destruction of these symmetries by displacement of the
Ta ion. 

\subsection{thermoelectric properties}

Additionally, thermoelectric parameters of \ttn~ are calculated by a constant scattering 
time approximation $\tau$. The results are shown in Fig. \ref{sf}. 
Note that these include only electronic contributions. 
Thus, the figure of merit $zT = S^{2}\sigma(E,T)/\kappa_{el}(E,T)$, given here,
is an upper bound.

\begin{figure*}[tbp]
\resizebox{16cm}{10cm}{\includegraphics{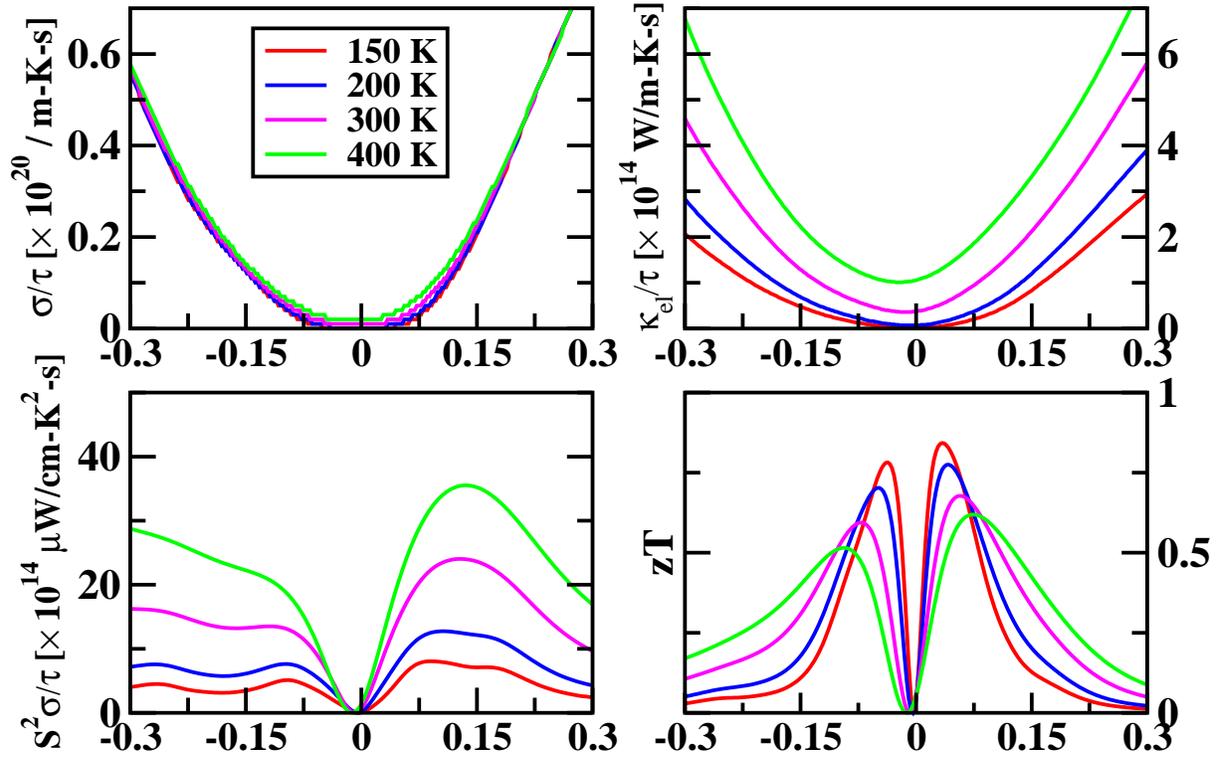}}
\caption{(Color online) (a) Electric $\sigma$ and (b) electronic thermal $\kappa_{el}$ conductivities, divided by the scattering time $\tau = 0.8 \times 10^{-14}$ sec. 
(c) Power factor and (d) figure of merit, contributed by electrons.
}
\label{sfig3}
\end{figure*}


\begin{thebibliography}{10}
\bibitem{gap} Using three different codes {\sc wien2k, fplo}, and {\sc vasp},\cite{vasp} 
our careful calculations with a dense $k$ mesh indicate that \ttn~ is either metallic 
or insulating with a tiny gap of several meV within the GGA, difference that are typical
between different codes.
Our value of the gap is different from the previous reports in Refs. \cite{shein,matar,polfus} online which quote energy gaps of 0.4 --0.8 eV.
However, one of them included an artficial factor to enhance the gap.\cite{shein}
The largest value was also obatined with a hybrid functional.\cite{polfus}
Thus the main difference in the quated gap among these calculations is due to 
using different exchange-correlation functionals or different basis sizes.
Most importantly, however, is that the previous calculations did not including spin-orbit
coupling. 
The lowest conducting bands from Ta $t_{2g}$ states undergo a substantial spin-orbit splitting,
which reduces and inverts the gap; see Fig. \ref{band}.
\end{thebibliography}
\end{document}